\documentstyle[prl,aps,epsfig]{revtex}
\begin{document}

 \newcommand \be {\begin{equation}}
\newcommand \ee {\end{equation}}
 \newcommand \ba {\begin{eqnarray}}
\newcommand \ea {\end{eqnarray}}
 \newcommand \bea {\begin{eqnarray}}
\newcommand \eea {\end{eqnarray}}
\def \(({\left(}
\def \)){\right)}

 \def \vn{{\mathbf{n}}}
 \def \vt{{\mathbf{t}}}
\def \vF{{\mathbf{F}}}
\def \vz{{\mathbf{z}}}
\def\Eel{E_{el}}
\def\EE{{\cal E}}
\def \eps{\epsilon}  
\def\bi{\bibitem}

\twocolumn[\hsize\textwidth\columnwidth\hsize\csname@twocolumnfalse\endcsname

\title{\bf  Elasticity model of a  supercoiled  DNA  molecule}

  \author{C. Bouchiat and M. M\'ezard}

\address{ Laboratoire de Physique Th\'eorique de l'Ecole Normale Sup\'erieure
 \cite{add}
  \\
24 rue Lhomond, 75231 Paris Cedex 05, France }

\maketitle

\begin{abstract}
Within a simple  elastic theory, we study  the elongation versus force
characteristics
 of a supercoiled DNA molecule at thermal equilibrium in the regime of small
supercoiling. The partition function
is mapped to the path integral representation for a quantum
charged particle in the  field of a magnetic monopole with unquantized charge.
 We show that the theory is singular in the continuum limit
and must be regularised at an intermediate length scale. We find good
agreement with existing experimental data, and point out how to measure the
twist rigidity  accurately.  LPTENS 97/28.
\end{abstract}

\twocolumn
\vskip.5pc]
\narrowtext

The measurements on single DNA molecules,
beside  their possible biological interest,  provide a wonderful laboratory for
the physical  studies of a single polymer chain. For
instance, recent experiments have shown that the elongation versus force
characteristics of a single DNA molecule \cite{smith}
 is very well fitted \cite{marsig} by the
well known worm-like chain (WLC) \cite{fixman} which  describes a chain by an
elastic
continuous curve at thermal equilibrium, with a single elastic constant, the
persistence length $A$ characterizing the bending energy.
The WLC can be solved analytically by mapping it
to a quantum mechanical problem. Its partition function is
nothing but a Euclidean path integral for a  quantum dumbbell,
which can be computed, in the relevant limit of long chains,
by finding the ground state of the  corresponding Hamiltonian.

Our work is motivated by the more recent experiments which have measured the
 elongation versus force
characteristics of a supercoiled DNA molecule \cite{strick}.
 We use the simplest generalization of the WLC with twist rigidity,
which is supposed to work at small supercoiling angles. 
The  DNA molecule is  described by
  a thin  elastic rod  involving a new elastic constant, the twist rigidity $C$.
  The stress imposed by twisting the endpoint of an open rod can be
absorbed both in some twist (if the axis of the rod is a straight
 line), and in some deformation of the rod's axis. This decomposition is
well known in the case of closed rods, for which the experimental
constraint is a topological invariant \cite{geometry}. It results in a
subtle competition involving the creation of plectonemes which has received
quite a lot of attention, both for the study of the ground state,
 and also taking care phenomenologically
 of thermal fluctuations around
some low energy configurations \cite{boles,marsig2,fain}.
In contrast, we keep here to the simplest regime of small
 supercoiling, but we provide a full analytic and numerical
study of the twisted open rod at thermal equilibrium at a finite temperature,
extending  thus the  standard WLC analysis to this case.
While it has been known for some time that the elastic thin rod
model agrees qualitatively with experimental results on DNA,
 this knowledge was based only on
some Monte Carlo simulations \cite{volog,marvol}. Our analytic
study of the  worm-like rod chain (WLRC) shows very good quantitative agreement
with the experimental data at small supercoiling and allows for
a precise determination of the ratio $C/A$, which was poorly known so far.

 We shall show
that beside the two elastic constants describing the bending and twisting
rigidity, one  needs to introduce an intermediate
lengthscale  $b$ between the microscopic interbasepair distance and
the persistence length, which  plays the role  
of a short distance cutoff. Obviously 
 the rod description cannot hold below a cutoff  length scale $b$ 
which is at least  of the order of the   double helix  pitch 
$ p\simeq 0.34 \ nm $, since our description averages over these periodic
oscillations.  As we shall see
  the WLRC continuous limit  is
singular and shows  properties qualitatively very different
 from any  discretized version of
the chain. So the existence of a cutoff, which is  expected on physical ground,
is crucial, but the final results turn out to
be fairly independent of its precise value, within a reasonable range around $ . 5 \  nm $.
 Similar singularities of the
continuous limit are well known in the winding properties of
pure random walks \cite{rw}. Their appearance here is  not
fortuitous since the  worm-like rod chain is related to random walks
in rotation space. It is interesting that supercoiled DNA presents
an experimental system where these subtleties of the continuous
limit of random walks turn out to be relevant.

 The  WLRC, already studied in \cite{volog,marsig2,fain,marvol},
 is described in the continuous limit
 by  the orthonormal triedron  $ \{ {\mathbf{t}}(s), {\mathbf{u}}(s),
{\mathbf{n}}(s)  \}$
where $s$ is the arc length along the molecule, $ {\mathbf{t}}$ is the unit
vector
tangent to the chain, and ${\mathbf{n}}$ describes the orientation of the
rod.
For describing DNA, this triedron
is obtained by applying a rotation $ {\cal R }(s) $ to a reference triedron
 which characterizes the natural helical structure
of the molecule. The rotation  $ {\cal R }(s) $ is parametrized
 by  the usual three  Euler angles $ \theta(s), \phi(s),\psi(s) $,
 and the  reference triedron  is such that $\theta(s)=0$,
$\phi(s)+\psi(s)=\omega_{0} s$, where
$ \omega_{0}$ is the rotation per unit length of the base axis in
a relaxed rectilinear DNA molecule.
 With the above  definition, the set of $s$ dependent Euler angles
 $ \theta(s), \phi(s),\psi(s) $ describes the general deformations
 of the DNA molecule with respect to the relaxed rectilinear configuration.
 We neglect self-avoidance, the energy of a
chain of length $L$ is purely elastic: $\Eel=\int_{0}^{L} ds (e_b+e_t)$,
 with the bend and twist energy densities given by:
 \bea
 e_b&=&{A \over 2} \vert  {d \vt(s) \over ds} \vert ^2= {A \over 2} \left(
{\dot{\phi}}^2 \,{\sin^2\theta}+{ \dot{\theta}}^2 \right) \nonumber \\
e_t&=&{C \over 2} {\vert  \vt(s) \times \vn(s) . {d \vn(s) \over d
s}-\omega_{0}\vert}^2 =
  { C \over 2} ( \dot{\psi} +\dot{\phi} \, \cos\theta )^2
\label{elast}
 \eea
where  the dot
stands for the  $s$ derivative. We  work in units where the
temperature $k_B T=1$, so that $A$ and $C$ have dimension of a length.
 The discretised version is
defined by quantifying $s$ as an integer multiple of
 an elementary length scale $b$, and approximating integrals and derivatives
by sums and differences, while keeping the periodicity.
We study the equilibrium properties of such a rod pulled by a force $\vF=F
\vz$. The total energy is thus $E=\Eel-F \int_0^L ds \cos\theta(s)$.

 The partition function of the elastic chain described by eq.(\ref{elast}) is
nothing but the Euclidean path integral for a quantum symmetric top,
with the important difference that
  the  eigenfunctions  are not  periodic in the angles
 $\psi $ and $\phi$. Therefore
the momenta conjugate to these angles will not be quantized.
 In our analytical work,  we  suppose for simplicity that
the boundary values of the Euler angles are $\theta(0)=\theta(L)=0$,
and we   define $\phi(0)=\psi(0)=0$. Then the experimentally imposed
supercoiling angle $\chi$ amounts to fixing:
$\psi(L)+\phi(L)=\chi $.  In the open DNA chain the continuous angle 
$\chi $  replaces the topological  linking  number $ L _k $ .  
We shall limit  ourselves to  configurations where the Euler angles are
regular enough, such  that $\chi $   can be
written as the integral $ \int_0^L ds \ \left( \dot{\psi} +\dot{\phi}\right)$. 
It is  convenient to introduce the rod twist $T_w$, given by the integral
  $T_w=\int_{0}^{L} ds  \left( \dot{\psi} +\dot{\phi} \, \cos\theta\right) $;  $T_w$ 
 appears as a Gaussian variable in the partition function path integral. We now
{\it{define} } a `local writhe' contribution to the supercoiling angle $\chi_W $ as
$\chi_W=\chi-T_w= \int_{0}^{L} ds \ \dot{\phi} (1-\cos\theta)$. The above 
decomposition \cite{fain} is reminiscent of the one performed in the case of of
closed chain where the topological linking number $ L_k$ is decomposed
into  twist  and writhe \cite{fn1}.

The partition function for a fixed value of
$ \chi $  is given by the  path integral in the space of Euler angles:
\be
Z=\int d[\cos\theta,\phi,\psi] \
 \delta \((\chi-\int_0^L ds (\dot{\phi}+\dot{\psi})\))
e^{- E}
\label{ZZ1}
\ee
 After introducing an integral representation of the $\delta$ function which
fixes $\chi$,
one can perform the gaussian path integral on the angle $\psi$. $Z$
is then expressed as a path integral on the two angles $\cos \theta$ and
$\phi$,
with an effective energy:
\be
{\cal E}= \int_0^L ds
(e_b -  F \cos\theta)
+{ C \over 2 L } \((\chi-\chi_W\))^2
\label{ZZ2}
\ee

 This form (\ref{ZZ2}) is useful for numerical simulations \cite{volog}
 after a proper discretization,
 but not  for analytic computation, due to its non local character.
 Alternatively we can compute   the $\chi$
Fourier transform  $ \tilde Z   =  \int d \chi  \ Z \exp(-i k \chi)$,
which is again given by  a path integral on the two angles $\cos \theta$ and
$\phi$,
with the effective energy:
\be
\tilde{\cal E}(k)=
{k^2 L \over 2 C}+\int_0^L ds
\left(e_b -  F \cos\theta
+i k  \dot{\phi} (1-\cos\theta) \right)
 \label{Ztf1}
\ee
 This last form has an appealing quantum mechanical interpretation:
If one analytically continues  the s-integral  towards the imaginary axis,
  one recognizes the action integral of a particle
 with unit charge
 moving  on the unit sphere under the joint action of the electric field $ F$
and the  magnetic field
 $ A_{\phi} = k\,(1-\cos\theta) $ of a magnetic monopole of charge k. One
easily deduces
the corresponding Hamiltonian $H$, by substituting  $
p_{\phi}=-i\frac{\partial}{\partial\,\phi }$
by $p_{\phi} -A_{\phi} $ in the WLC
  Hamiltonian (which corresponds to $ A_{\phi}=0$).
Because of the averaging over the final $ \phi = \phi(L)$,
 only the eigenvalue $m=0$ of $ p_{\phi} $
  contributes and  we can set $ p_{\phi}=0 $ in $H$. We  work with the
dimensionless quantities $\hat H=H/A$ and $\alpha \equiv  A F $, in terms of
which we get:
\be
\hat H=
-\frac{1}{ 2\,\sin\theta }\frac{\partial}{\partial\,\theta}\sin\theta \,
  \frac{\partial}{\partial \theta }-
\alpha \cos\theta+{k^2 \over 2} {1-\cos\theta \over 1+\cos\theta}
 \ee

  Introducing  the eigenstates and the eigenvalues of $\hat H$ ,
$ \hat H{\Psi}_{n}(k,\theta)= {\epsilon}_n(k^2,\alpha) {\Psi}_{n}(k,\theta)$,
 the Fourier transformed partition function $\tilde Z$ can be written as the
sum:
\be
\tilde Z=\sum_n {\vert {\Psi}_{n}(k,0)\vert}^2
\exp\((- \frac{L}{A} \(({\epsilon}_n(k^2,\alpha)+ {k^2 A \over 2 C }\))\))
\ee
In the large $L$ limit, the sum over the eigenstates is dominated by the one
 with lowest energy $\eps_0(\alpha,k^2)$, if $ L/A \gg \Delta\eps $
where $\Delta\eps$ is the energy
gap between the ground state and the nearest excited state of  $\hat H$.
  This gives the approximate expression for the partition function
$ Z$ :
\be
Z  \simeq \int dk \  \exp\((- \frac{L}{A} \((  {\epsilon}_0(k^2,\alpha) +
{k^2  A \over 2 C} \))+i\, k\,\chi  \))
\label{Zcol}
\ee
Therefore one can deduce, from the  ground state energy $\eps_0(\alpha, k^2)$
of
the Hamiltonian $\hat H$,  the observable properties of a long WLRC, of
which we now discuss two important ones.
The relative extension
of the  chain in the direction of the force is given by
${\langle z \rangle / L} = (A / L) {\partial \ln Z \over \partial \alpha}$.
 If instead of constraining $\chi$
 one  measures its thermal fluctuations,
their probability distribution is just $P(\chi) \propto Z$. For
instance  the
second moment is given by:
\be
   <\, {\chi}^2 \, > = { L\over C}+
  { 2\, L\over A} \lim_{k^2\to 0}{ \partial \epsilon_0(k^2,\alpha ) \over
\partial\, k^2 }
\label{chi2}
\ee

This expression  shows that the WLRC is pathological because of "giant" 
fluctuations of $\chi_W$.
The contribution to $<\chi^2>$ from the twist fluctuations, ${ L\over C}$,
scales
linearly in $L$,
as one expects in a one dimension  statistical
mechanics system with a finite correlation length.
In contrast the second piece of (\ref{chi2}) giving the contribution 
from $\langle \chi_W^2 \rangle$, is divergent: evaluating
$\epsilon_0(k^2,\alpha )$ at
small $k^2$ from standard perturbation theory, we find
$ \langle{\chi_W}^2 \rangle
    =  (L/A) \langle\,(1-\cos\theta)/( 1+\cos\theta)\,\rangle_0
$
where $  {\langle \; \;\rangle}_0 $ is  the quantum average taken on the
groundstate
 $\Phi_0(\theta)$
   of the WLC Hamiltonian (which is $\hat H$ at $ k=0$). As
 $\Phi_0(\pi)\ne 0$ (for any finite force), we get  a logarithmically divergent
result.
One can show that $\eps(\alpha,k^2)
\sim \eps(\alpha,0)+ |k| \Phi_0(\pi)^2$.  This linear behaviour
of the energy in $|k|$ shows that $P(\chi)$ has a Cauchy tail, and thus a
diverging
second moment. The Cauchy distribution has been verified in the limit 
of a vanishing force,
$\alpha=0$, where the eigenfunctions
can be found exactly in terms of Jacobi polynomials.
 A related consequence
is that
 the extensive part of the
average extension is unchanged by the supercoiling angle $\chi$ at small
forces:
$<z>/L\simeq 2 \alpha/3$ independently of $F$, in striking contradiction to
experiment.

In contrast to the WLC,  the continuous limit of the WLRC is singular.
This singular behaviour could have been anticipated since   $ \hat H $
describes the motion of a charged particle in a  magnetic monopole
with an {\it{ unquantized}}  magnetic
charge,  a  notoriously ill defined problem if no cutoff is provided near
$\theta = \pi$.
As we have argued  previously, the  WLRC model is not expected to give
a good description of supercoiled DNA unless   one introduces the cut-off
length scale $b$.
   The non trivial fact is that this existence of
a cutoff affects the `macroscopic' properties taking place on the length
scale of the whole molecule.

In order to validate the discretized   WLRC,
we have performed a Monte Carlo simulation,
mostly using the discretized version of (\ref{ZZ2}). Such simulations
are known to account  well for
the observed behaviour of circular DNA \cite{volog}, and have been
used recently for the study of chains elongated with large
supercoiling angles \cite{marvol}. With respect to these works, we have
discarded the self avoidance, since our aim is to test the
discretized WLRC without self avoidance.
We have
discretized the chain with elementary rods of length $b=A/10$,
and simulated mostly chains of length $L=30 A$. In order to facilitate
the thermalisation, we have relaxed in the simulation the constraint
$\theta(L)=0$, which should not affect the extensive quantities. The
elementary moves which we used was to scan
sequentially each point $i=1,...,N=L/b$ in the chain, and propose
a global rotation of the tangent vectors $\vt_j,j=i,...,N$ around a random
axis with an angle $\gamma$ taken with
a flat distribution in $[-\gamma_0,\gamma_0]$, where $\gamma_0$ is chosen
such that the acceptance rate of the moves  is of order $.5$.
 These rotations of
a fraction of the chain were the best we found for insuring a relatively
fast thermalization \cite{BM2}.
 The results presented in Fig.1,
obtained with $C/A=1.4$, show that the elongation versus $\chi$
 characteristics
reproduces  well the experimental values at small enough $\chi$.

\begin{figure}
\centerline{\epsfxsize=7cm
\epsfysize=9cm
\epsffile{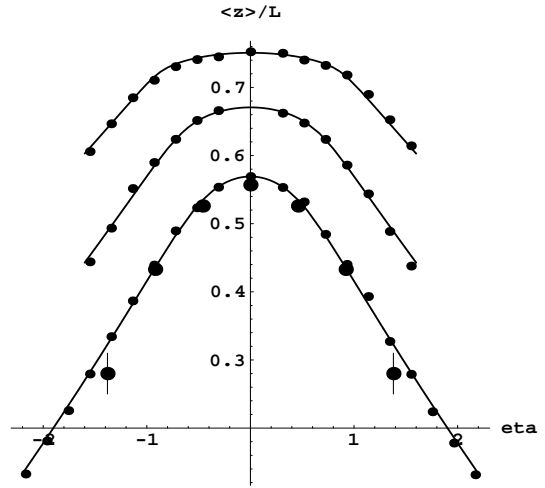}
}
\caption{The  elongation versus reduced supercoiling angle $\eta= \chi A/L $
for forces $ F=.116 ,.197,.328 pN $, from bottom to top.  $\chi$ is taken directly from experiment
and $\eta $ has been computed with $ A=51.35 nm $ and $ L=15.61\mu m $. The experimentalist's
$ \sigma  $ is related to $ \eta $ by $\eta=\omega_0 A \sigma=2\pi \frac{A}{p}=94.8 \sigma $. 
The smaller points are the experimental results, the bigger points on the lowest curve
 are from Monte Carlo simulations, the   full  lines is the analytic study 
of the WLRC through
the parametric representation (\ref{col}) with the values $ C/A $
given in the text. }
\label{fig1}
\end{figure}

The regularization in the discrete model comes from the fact that the
angle  $\phi-\phi'$ is defined modulo $2 \pi$. The computation of
$<\chi_W^2>$ can be done explicitely at $\chi=0$, and one finds that
the continuum expression $L/A \langle (1-\cos\theta)/(1+\cos\theta)
\rangle$
must be substituted
by $L/A \langle (1-\cos\theta)/(1+\cos\theta) R(\sin^2\theta A/b)\rangle$,
where the regularization function is given in terms
of Bessel functions by: $R(x)=I_1(x)/I_0(x)$.
 It is reasonable to assume that
the discrete model is well approximated by the Hamiltonian $ \hat H^{r}$
which is obtained from  $\hat H$ by this same substitution.
 We have computed  the
elongation properties of the WLRC
from the ground state energy of the corresponding regularized
 rod hamiltonian $ \hat H^r$. The variations of
$ {<z> \over L} $ with $ \chi $ now scale as  a function of $ \chi/L$,
as in experiments.  We introduce the intensive linking variable $ \eta= \chi
\; A/L $
(related to the experimentalists' $\sigma$ by $\sigma= \eta/\omega_0 A$).
The partition function in (\ref{Zcol}) can be computed by the saddle point
method in the limit
$ L/A\gg 1$ with $ \eta $ kept fixed. The saddle point is imaginary, $k=i
\kappa(\alpha)$,
and from its value one easily deduces the  elongation of the chain
using the general formulas given above. We  obtain  in this way the following
parametric
representation of  the bell shape curves giving $ {\langle z \rangle \over L} $
  versus
 $ \eta$, for a fixed value of the force $ \alpha $.
\be
 {A \over C} +2 {\partial \eps_0 \over \partial k^2}(\alpha,-\kappa^2)={\eta
\over \kappa}
\  ; \   {\langle z \rangle \over L}
=-{\partial \eps_0 \over \partial \alpha} (\alpha,-\kappa^2)
\label{col}
\ee
An alternative to solving the wave equation  associated with the  regularized 
hamiltonian   $ \hat H^r$ is to work  directly with the discretized
version of the effective energy (\ref{Ztf1}), using standard transfer matrix
methods. We have checked that these two procedures give results
for $\eps_0 (\alpha,-\kappa^2)$ which are in very good agreement. 
 Using interpolation techniques, one can
 eliminate  $ \kappa $ in (\ref{col} )  in order to obtain 
the ratio $ C/A $ in terms of $ \eta $, $ {\langle z \rangle / L} $
and $b/A$. For various values of the force and of $ b/A$,
 we have computed the
empirical value of $C/A$ obtained from each experimental point
 $ \eta $, $ {\langle z \rangle / L} $. 
With $b/A=.12$ all these values cluster nicely, allowing for a rather precise
 determination of $ C/A $. For three values of the force,
$F=.116 \; pN ; \ .197 \;pN  ; \ .328 \; pN$, we find respectively
 $C/A= 1.67 \pm .12 \ ; \ 1.66 \pm .10 \ ; \ 1.71 \pm .09$
(in this analysis we have restricted to the range of
small enough supercoiling: $  \vert \eta  \vert \le 1.5 $
 for $ F=.197 ,.328  \; pN $ and $  \vert \eta  \vert \le  2.2 $ for $F=.116 \; pN$ ).
 For each force, the  value  of $ C/A $ is the result of  a statistical  weighted 
average and the quoted  error is just the root mean square deviation,
read off directly  from  the data.
These three results agree and indicate a value of $ C/A\simeq 1.68 $.
However it should be stressed
that the experimental points used here are preliminary and the study 
of systematic errors in the experiment is not yet
completed so this value of $ C/A$ is  not yet definitive.
The  bell shape curves, computed for each  force with
the corresponding  value of  $ C/A$, are compared to the experimental points 
in fig1. The agreement 
looks  very satisfactory, indicating an overal consistency of the procedure.
  We have performed 
 the same analysis with  $ b/A = .05, .2 $.  With  $ b/A=.2 $
the  agreement remains  fairly good, but the value $ C/A $ for the 
 three values of the  forces exhibit fluctuations around 1.68 which can  reach $ 20 \% $.
 The value $ b/A= .05 $ appears to be excluded. 

Another  possible   method  to get the ratio $ C/A $  could be to measure
   the curvature of the bell shape curve , 
 $ \Gamma_{\eta}=\partial^2 (<z>/L)/\partial  \eta ^2$, at its maximum, which
    can be related to 
 $a_1(\alpha) = 2\,\(({ \partial \over \partial\, k^2
\,}{\epsilon}_0(k^2,\alpha )\))_{k=0} $ through perturbation theory. The
 $A/C$ is given in terms of the known function $a_1(\alpha)$ \cite{BM2} as:
\be
{A \over C}=-a_1(\alpha)+\sqrt{{\partial a_1/\partial \alpha \over \Gamma_\eta}}
\label{gamma}
\ee
At the moment the measurements of 
$ \Gamma_\eta  $ is however not precise enough for a good determination
with this method.

We have shown that the WLRC must be regularised at small length scale. The
corresponding
model can be solved analytically
and it accounts well for experimental results at small supercoiling, giving
a good method to determine the elastic constants ratio $C/A$.
Obviously our theory is limited to the small force - small
supercoiling regime. For instance the experiments show
that for $F >.45 pN$ the extension is not symmetric for $\chi \to - \chi$.
This kind of effect is totally beyond our simple elastic model
which is intrinsically symmetric. Extending it requires the introduction
of self avoidance to treat properly the plectoneme formation. The self avoidance will
also naturally provide a cutoff length scale. However taking it into
account properly is a major challenge \cite{moroz}.

We wish to thank J.-F. Allemand, D. Bensimon, V. Croquette
and T.R. Strick for numerous exchanges,
as well as J.-P. Bouchaud, A. Comtet and C. Monthus for useful
discussions on the winding distribution of random walks. While completing
this work, we learned that P. Nelson and his collaborators
 are working  on the same problem in
the limit of large forces.

\end{document}